\documentclass[sigconf]{acmart}
\pdfoutput=1
\usepackage{graphicx}
\usepackage{subfigure}
\usepackage{threeparttable}
\usepackage[normalem]{ulem}
\useunder{\uline}{\ul}{}
\usepackage{enumitem}
\allowdisplaybreaks[4]
\usepackage{amsmath}
\usepackage{makecell}
\usepackage{multirow}
\usepackage{graphicx}
\usepackage{tabularx}
\usepackage{xcolor}
\usepackage{algorithm} 
\usepackage{algpseudocode}

\usepackage{balance}

\AtBeginDocument{%
  \providecommand\BibTeX{{%
    \normalfont B\kern-0.5em{\scshape i\kern-0.25em b}\kern-0.8em\TeX}}}

\setcopyright{acmcopyright}
\copyrightyear{2022}
\acmYear{2022}

\acmBooktitle{Conference'17, July, 2017, Washington, DC, USA}

\makeatletter
\def\@ACM@checkaffil{
    \if@ACM@instpresent\else
    \ClassWarningNoLine{\@classname}{No institution present for an affiliation}%
    \fi
    \if@ACM@citypresent\else
    \ClassWarningNoLine{\@classname}{No city present for an affiliation}%
    \fi
    \if@ACM@countrypresent\else
        \ClassWarningNoLine{\@classname}{No country present for an affiliation}%
    \fi
}
\makeatother

\begin{document}

\title{Task Aligned Meta-learning based Augmented Graph for Cold-Start Recommendation}


\author{Yuxiang Shi$^{1}$, Yue Ding$^{1}$, Bo Chen$^{2}$, Yuyang Huang$^{1}$, Yule Wang$^{1}$, Ruiming Tang$^{2}$, Dong Wang$^{1}$}
\affiliation{
$^{1}$Shanghai Jiao Tong University, 
$^{2}$Noah's Ark Lab, Huawei
}



\begin{abstract}

The cold-start problem is a long-standing challenge in recommender systems due to the lack of user-item interactions, which significantly hurts the recommendation effect over new users and items. Recently, meta-learning based methods attempt to learn globally shared prior knowledge across all users, which can be rapidly adapted to new users and items with very few interactions. Though with significant performance improvement, the globally shared parameter may lead to local optimum. Besides, they are oblivious to the inherent information and feature interactions existing in the new users and items, which are critical in cold-start scenarios. In this paper, we propose a Task aligned Meta-learning based Augmented Graph (TMAG) to address cold-start recommendation. Specifically, a fine-grained task aligned constructor is proposed to cluster similar users and divide tasks for meta-learning, enabling consistent optimization direction. Besides, an augmented graph neural network with two graph enhanced approaches is designed to alleviate data sparsity and capture the high-order user-item interactions. We validate our approach on three real-world datasets in various cold-start scenarios, showing the superiority of TMAG over state-of-the-art methods for cold-start recommendation. 

\end{abstract}



\maketitle

\section{Introduction} 

Recommender systems aim to discover users' interests and have been extensively employed in various online systems \cite{wang2018billion, ying2018graph, chen2021airec}, such as E-commerce platforms, online advertising and social platforms. Despite the success of traditional matrix factorization (MF) models \cite{bokde2015matrix}, or popular deep learning models \cite{NCF, wang2024extraction}, one of the major challenges for most recommendation methods is the cold-start problem \cite{schein2002methods, gantner2010learning} because of the lack of user-item interactions. Since new users may abandon the system when they receive poor recommendations initially \cite{mcnee2006being}, it is critical to address this problem.

\begin{figure}[]
    \setlength{\abovecaptionskip}{-0.1cm}
    \setlength{\belowcaptionskip}{-0.2cm}
    \centering
    \centering
    \subfigcapskip=-8pt
    \subfigure[Conventional meta-learning framework]{
    \begin{minipage}[c]{0.5\textwidth}
    \centering
    \includegraphics[width=0.9\linewidth]{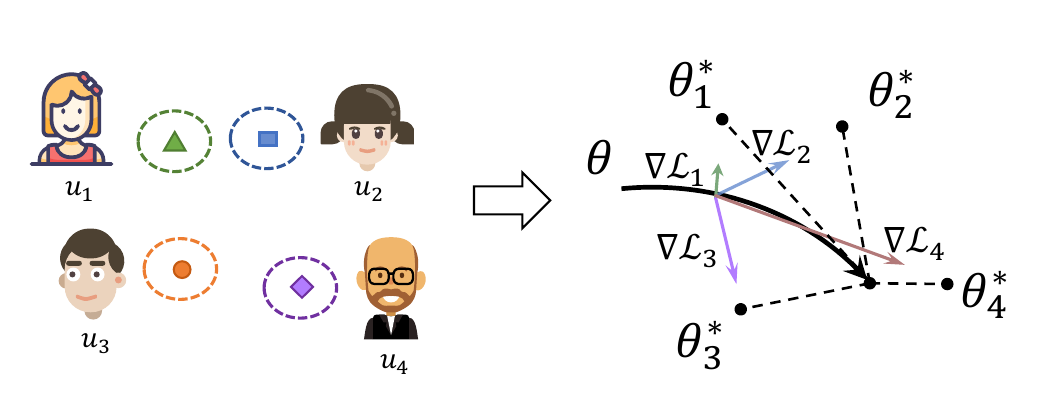}
    \label{fig:concept_a}
    \end{minipage}%
    }\vspace{-4mm}
    \subfigcapskip=-6pt
    \subfigure[Proposed TMAG framework]{
    \begin{minipage}[c]{0.5\textwidth}
    \centering
    \includegraphics[width=0.9\linewidth]{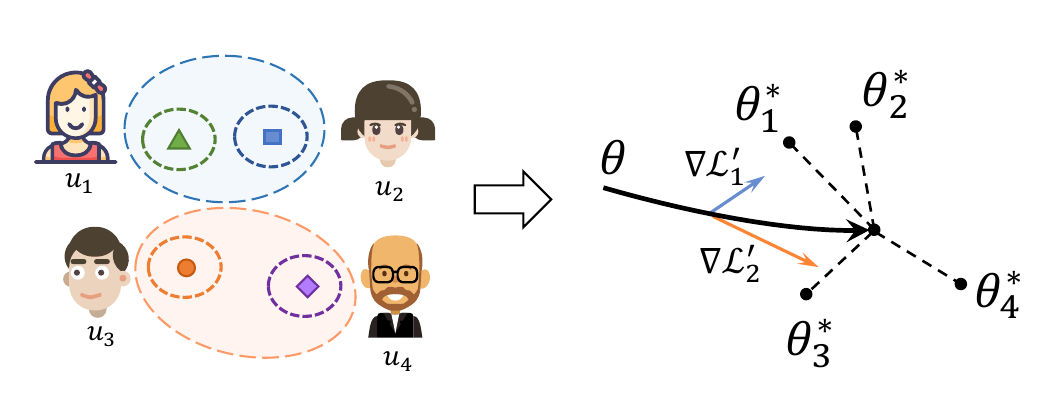}
    \label{fig:concept_b}
    \end{minipage}%
    }
    \caption{Visualization of model parameter $\boldsymbol{\theta}$ of the conventional framework with 4 tasks and the proposed TMAG with 2 tasks aligned by ages. The gradient descent direction in conventional meta-learning framework is biased towards $u_4$, while in TMAG users in the same group have consistent optimization direction thus avoiding such local optimum.}
    \label{fig:concept}
\end{figure}




Traditional way to alleviate the cold-start problem is utilizing \textbf{feature-level} strategies, which can be categorized into two groups. Firstly, modeling \textit{inherent information}, (e.g., user profiles \cite{BriandSBMT21, gantner2010learning}, item attributes \cite{roy2016latent, chen2020esam} and cross-domain knowledge \cite{hu2018conet, krishnan2020transfer}) to enhance the representations of new users or items. 
Secondly, modeling \textit{feature interaction} via graph neural networks (GNN) \cite{WangZWMHW21} and heterogeneous information networks (HIN) \cite{liu2020heterogeneous, MvDGAE} to capture the high-order collaborative signal.
Despite progress, these method handle the cold-start issue from the feature-level, which heavily relies on the availability and quality of features.

 
On another line, at the \textbf{model-level}, recent works in few-shot learning \cite{vinyals2016matching} and meta-learning \cite{vanschoren2018meta} have made prominent progress to solve the data sparsity problem in various fields. Most of the current meta-learning methods \cite{MeLU, wei2020fast, MetaHIN} adopt optimization-based algorithms (e.g., MAML \cite{finn2017model}) to address the cold-start problem. 
The main idea is to learn a global parameter to initialize the parameter of personalized models. These methods construct diverse few-shot user preference tasks that mimic the cold-start scenarios and extract meta-knowledge across meta-training tasks as a strong generalization prior. Then the learned prior knowledge can be rapidly adapted to new users with scarce interactions during meta-testing. Practically, they have achieved promising results in cold-start recommendation. 


However, the existing meta-learning methods have the following limitations. They formulate each user as a task and learn globally shared meta-knowledge across all users. The coarse-grained global knowledge leads the model to local optimum when dealing with users whose gradient descent directions are different from major users \cite{dong2020mamo}. As illustrated in Figure \ref{fig:concept_a}, $\nabla\mathcal{L}_4$ dominates the direction of gradient descent due to the age difference. Therefore, the parameter $\boldsymbol{\theta}$ is biased towards the optimal solution. In addition, existing methods are deficient in taking full advantages of both inherent information and feature interactions, which is critical to model new user and item. 
These observations raise two research challenges: 
\begin{itemize}[leftmargin=*]
\item How to ease the local optimum of meta-learning in cold-start scenarios?
\item How to make full use of both inherent information and feature interactions to alleviate the sparsity?
\end{itemize}

With these challenges in mind, in this paper, we propose a $\textbf{T}$ask aligned $\textbf{M}$eta-learning based $\textbf{A}$ugmented $\textbf{G}$raph (TMAG) approach to address the cold-start recommendation at \textbf{both feature-level and model-level}. For the first challenge, a fine-grained task aligned constructor is proposed to cluster similar users and divide tasks for meta-learning. Specifically, an attribute-oriented autoencoder is used to extract latent representations for users and items according to the inherent attributes. Then, users with similar representations are clustered into a group, which is regarded as a task and has consistent optimization direction, thus easing the local optimum during the meta-training. For the second challenge, we propose an augmented graph neural network to capture the high-order user-item interactions. Specifically, two graph enhanced approaches are utilized to alleviate the data sparsity and explore potential interactive signals from the perspective of attribute and graph structure, respectively.

The major contributions of our work are summarized as follows:

\begin{itemize}[leftmargin=*]
\item We propose a fine-grained task aligned constructor to capture the latent clustering knowledge that can be rapidly adapted to new users, which can address the local optimum issues. 

\item We augment the adjacency matrix of the graph by combining the graph structure information and attribute information, which alleviates the data sparsity problem.

\item We employ a task-wise attribute contrastive regularization to enhance the latent clustering knowledge.

\item We conduct extensive experiments on three public real-world datasets in various cold-start scenarios to demonstrate the state-of-the-art performance of TMAG. 
 
\end{itemize}

\section{Related Work}

\subsection{Cold-start Recommendation}
A prevalent strategy to address the cold-start problem relies on side information, by incorporating user profiles or item content \cite{volkovs2017dropoutnet, van2013deep, BriandSBMT21, gantner2010learning} into traditional collaborative filtering \cite{ItemCF, linden2003amazon, NCF}. In particular, many approaches focus on clustering warm users by modeling group-level behaviors and then assigning cold users to existing clusters using the side information \cite{wu2016cccf, ma2019dbrec, krishnan2018insights, hu2018conet}. Building on these works, we will also utilize side information and incorporate a clustering component in Section \ref{method}. Beyond these content-based features and user-item interactions, another strategy is to augment data via adversarial regularization \cite{krishnan2018adversarial, chae2019rating, chae2020ar}, but adversarial learning is computationally intensive and cannot handle large massive. Some transfer learning-based methods \cite{man2017cross, bi2020dcdir, krishnan2020transfer} alleviate the cold-start problem by transferring well-learned representations of overlapped objects from the source domain to the target domain. The active learning scheme \cite{zhu2020addressing, jacobson2016music, shi2017local} explicitly encourages new users to rate items from the catalog through various interview processes with the extra costs or budgets. Additionally, GNN-based models \cite{Pinsage, NGCF, LightGCN, WangZWMHW21, wang2021deminet, li2021extracting, wang2021icmt} have been developed for recommendation, which utilizes user-item bipartite graph to capture high-order collaborative signal. In general, GNN-based methods maximize a user's likelihood of adopting an item, rather than directly improving the embedding quality of cold-start users or items. We will improve them in our method to do fast adaption in cold-start scenarios. While these solutions demonstrate promising performance, they only address the cold-start issue at the feature-level and heavily rely on the availability and quality of side information or extra domains.

\subsection{Meta-learning for Recommendation}
Known as learning to learn, meta-learning mines the underlying patterns behind user-item interactions. Meta-learning aims to extract meta-knowledge across tasks that can be rapidly adapted to a new task with limited examples.
Previous works on meta-learning for cold-start recommendation can be classified into two categories. One is the metric-based method \cite{vartak2017meta, sankar2021protocf}, which learns a similarity metric between new instances and examples in the training set. ProtoCF \cite{sankar2021protocf} learns to compose robust prototype representations for few-shot items without using side information. Most of meta-learning methods are optimization-based and introduce the framework of MAML \cite{finn2017model} into cold-start recommendation \cite{song2021cbml, yu2021personalized, zheng2021cold, du2019sequential}. 
MeLU \cite{MeLU} divides the model parameters into personalized parameters and the embedding parameters. These parameters are updated in the inner loop and outer loop respectively.
MetaHIN \cite{MetaHIN} applies heterogeneous information networks (HIN) to capture rich semantics of meta-paths. 
MetaCF \cite{wei2020fast} learns an accurate collaborative filtering model well-generalized for fast adaption on new users.
The above works learn globally shared parameters that are the same for all tasks. This may
lead the model to the local optimum and converge slowly. To address this problem,
MAMO \cite{dong2020mamo} improves MeLU by introducing task-specific memories and feature-specific memories to provide personalized bias terms when initializing the parameters.
PAML \cite{wang2021preference} improving existing framework with better generalization capacity by leveraging users’ social relations.
In contrast to existing works, to solve above issue, we cluster users with similar attributes into the same task to capture latent clustering knowledge that can be rapidly adapted to new users.

\begin{figure*}[t]
    \setlength{\abovecaptionskip}{-0.1cm}
    \setlength{\belowcaptionskip}{-0.1cm}
    \centering
    \includegraphics[width=1.0\linewidth]{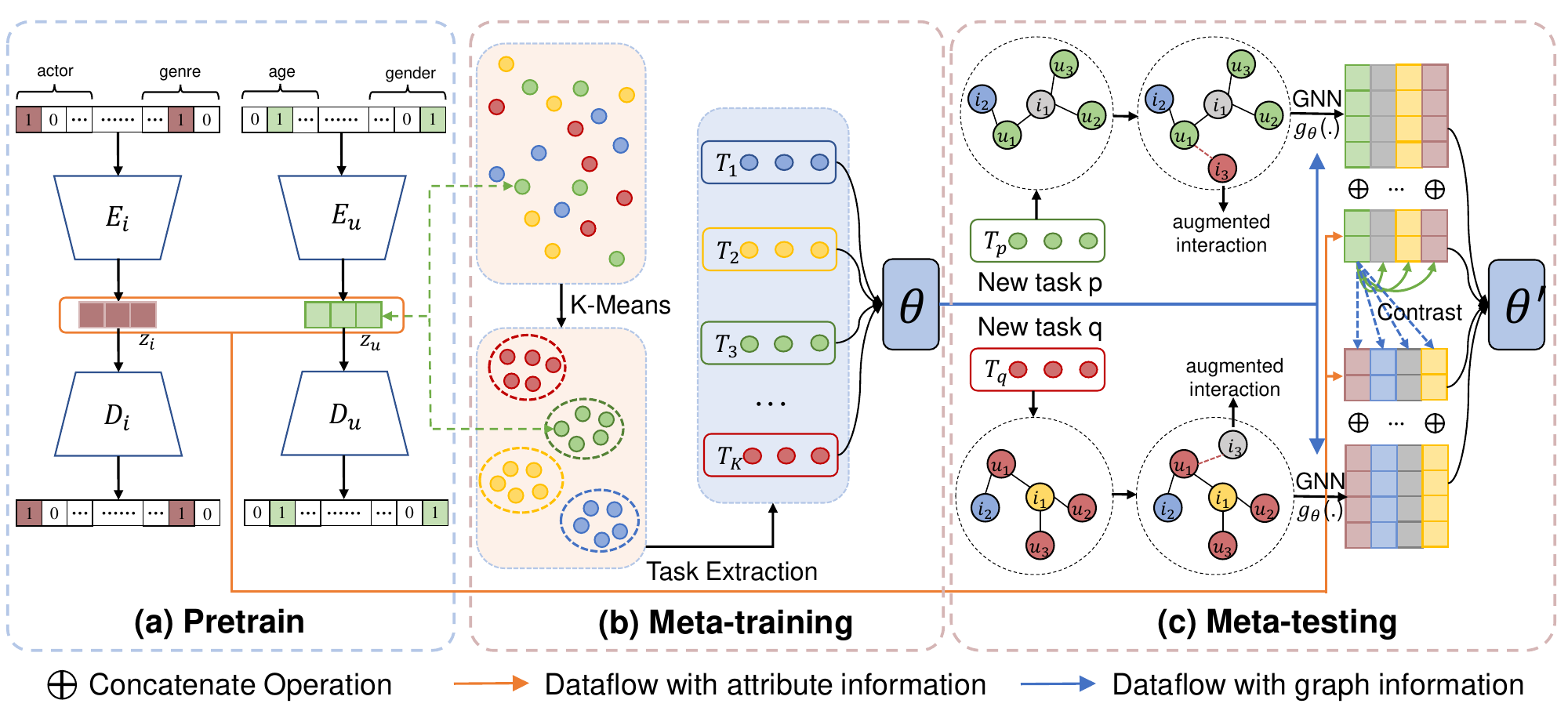}
   \caption{The overview of the proposed TMAG framework. (a) In the pretrain phase, we use two attribute-oriented autoencoders to learn user and item attribute embeddings. (b) In the meta-training phase, we cluster users into different groups and formulate each group of users as a task. These tasks are leveraged to train the meta-training model, and we obtain the parameter $\boldsymbol{\theta}$. (c) In the meta-testing phase, firstly, we employ the learned $\boldsymbol{\theta}$ as the initialization parameter to get user and item representations. Then, we augment the adjacency matrix of the graph by combining the graph structure information and attribute information. At last, we employ the updated $\boldsymbol{\theta}'$ for recommendation.}
    \label{fig:framework}
\end{figure*}

\section{Preliminary}
\subsection{Problem Definition}
Let $\mathcal{U}$ = $\left\{u_{1}, u_{2},...,u_{M}\right\}$ and $\mathcal{I}$ = $\left\{i_{1}, i_{2},...,i_{N}\right\}$, where $\mathcal{U}$ denotes the user set, $\mathcal{I}$ denotes the item set, $M$ and $N$ denote the number of users and items respectively.  Therefore, we use a matrix $\boldsymbol{R} \in \mathbb{R}^{M \times N}$ to denote the interaction behavior between users and items. Formally, 
\begin{equation}
    \boldsymbol{R}_{ui}=
    \begin{cases}
    1,& \text{if user $u$ has interacted with item $i$;}\\
    0,& \text{otherwise.}
    \end{cases}
  \label{eq:R}
\end{equation}
The user-item interactions can be transformed to a user-item bipartite graph $\mathcal{G} = (\mathcal{V}, \mathcal{E})$, where $\mathcal{V}$ denotes the node set and $\mathcal{E}$ denotes the edge set in the graph, and we have $|\mathcal{V}| = (M + N)$ denoting the number of nodes in the graph. $\boldsymbol{A} \in \mathbb{R}^{M \times N}$ is the adjacency matrix of $\mathcal{G}$. 


Cold-start issue is a fundamental challenge in recommender systems, which can be categorized into complete cold-start and incomplete cold-start \cite{wei2020fast}. In the former, there are no interactions for new users and in the latter, there are only a few interactions for new users. In this paper, we are interested in the incomplete cold-start problem. Our model serves as an additional cold-start recall method. To evaluate the model performance, we divide the recommendation task into three sub-tasks as follows,

\noindent$\boldsymbol{Task1:}$ Recommend existing (old) items to new users;

\noindent$\boldsymbol{Task2:}$ Recommend new items to existing (old) users;

\noindent$\boldsymbol{Task3:}$ Recommend new items to new users.


\subsection{Background on Meta-learning}
Meta-learning learns generic knowledge from a large class of tasks and generalizes it to new tasks. In meta-learning, we have a task set $\mathcal{T}$ which consists of meta-training tasks $\mathcal{T}^{tr}$ and meta-testing tasks $\mathcal{T}^{te}$. And we define two sets, the support set $\mathcal{S}_k$ and query set $\mathcal{Q}_k$ for each task $\boldsymbol{T}_k \sim p(\mathcal{T})$. Optimization-based meta-learning methods attempt to find desirable parameter $\boldsymbol{\theta}$ of the model $f$. During meta-training, there are two rounds of updates for $T_k \in \mathcal{T}^{tr}$. In the inner-loop update, the model adjusts the parameter $\boldsymbol{\theta}$ to $\boldsymbol{\theta}_k$ by gradient $\nabla_{\boldsymbol{\theta}}\mathcal{L}_k(f_{\boldsymbol{\theta}})$ in $\mathcal{S}_k^{tr}$, where $\mathcal{L}_k(f_{\boldsymbol{\theta}})$ denotes the training loss on task $T_k$ with global parameter $\boldsymbol{\theta}$. In the outer-loop update, the model trains the parameter to minimize the loss on the query set $\mathcal{L}_k(f_{\boldsymbol{\theta}_k})$ in $\mathcal{Q}_k^{tr}$, where $\mathcal{L}_k(f_{\boldsymbol{\theta}_k})$ is the testing loss on task $T_k$ with task-specific parameter $\boldsymbol{\theta}_k$. During the meta-testing process, the model only updates the parameter in the inner-loop on the support set $\mathcal{S}^{te}$ and evaluate on the query set $\mathcal{Q}^{te}$.

\section{METHODOLOGY}
\label{method}
\subsection{Overview of TMAG}
The framework of TMAG is illustrated in Figure \ref{fig:framework}. There are three components in the framework: (1) a task aligned constructor that extracts attribute embeddings and constructs tasks to capture latent clustering knowledge; (2) an augmented graph neural network to alleviate data sparsity and capture the high-order user-item interactions; and (3) a contrastive regularization that enhances the latent clustering prior knowledge.

\subsection{Task Aligned Constructor}
\subsubsection{Attribute-Oriented autoencoder}
The proposed attribute-oriented autoencoder is shown in Figure \ref{fig:framework} (a). The model takes the user content (e.g., age and gender) and the item content(e.g., actor and genre) as the input. We employ an attribute-oriented autoencoder to obtain the user and item feature embeddings. First, we project the input into a low-dimensional hidden embedding. Second, from the latent embedding, we reconstruct the input content in the output space. 

Most attributes are categorical features. We define the user content input $\boldsymbol{x}_u$ for user $u$ and item content input $\boldsymbol{x}_i$ for item $i$ as follows, 
\begin{equation}
\boldsymbol{x}_u = [c_{u1};\cdots;c_{uP}]; \quad \boldsymbol{x}_i = [c_{i1};\cdots;c_{iQ}],
\end{equation}
where $P$ is the number of user content field, $c_{up}$ is $d_p$-dimensional one-hot or multi-hot vector for content $p \in \left\{1,\cdots,P \right\} $ of user $u$. Similarly, $Q$ is the number of item content field, $c_{iq}$ is $d_q$-dimensional one-hot or multi-hot vector for content $q \in \left\{1,\cdots,Q \right\} $ of item $i$.

We choose Rectified Linear Unit (ReLU) as the non-linear activation function to obtain the latent user embedding as follows, 
\begin{equation}
\boldsymbol{z}_u = \sigma (\boldsymbol{W}_u^1 \boldsymbol{x}_u + \boldsymbol{b}_u^1),
\end{equation} 
where $\boldsymbol{W}_u^1, \boldsymbol{b}_u^1$ are trainable parameters. Here we do not utilize graph convolutional network (GCN) \cite{GCN} to encode the user content because we note that aggregating too much information from neighbors affects the expressive power of user's attributes for the aligned task in meta-learning. And we design a decoder to reconstruct $\boldsymbol{x}_u$ as follows,
\begin{equation}
\boldsymbol{x}_u^r = \sigma (\boldsymbol{W}_u^2 \boldsymbol{z}_u  + \boldsymbol{b}_u^2),
\end{equation} 
where $\boldsymbol{W}_u^2, \boldsymbol{b}_u^2$ are trainable parameters.
Mathematically, the objective function of the user attribute-oriented autoencoder is, 
\begin{equation}
\boldsymbol{L}_u = \sum_{u \in \mathcal{U}} \left \| \boldsymbol{x}_u - \boldsymbol{x}_u^r \right \|^2 + \lambda \left \| \boldsymbol{W}_u \right \|^2,
\label{auto_loss}
\end{equation} 
where $\boldsymbol{W}_u$ denotes all trainable model parameters; $\lambda$ controls the $L_2$ regularization strength to prevent over-fitting. Analogously, we can obtain the latent item embedding $\boldsymbol{z}_i$ utilizing a new item attribute-oriented antoencoder. 

\subsubsection{Task Construction}
In incomplete cold-start recommendation scenarios, users only have a few interactions with items. Therefore, user and item representations learned from user-item interaction pairs are inadequate. When new users (items) arrive, we are supposed to capture latent clustering knowledge among users with similar attributes to obtain more expressive representations. 
Similar to the classical meta-learning method MAML, we construct various meta tasks to locally share latent clustering knowledge. To be specific, we utilize the K-Means algorithm \cite{likas2003global} to divide all users $\mathcal{U}$ into different $K$ task groups according to $\boldsymbol{x_u}$ ($u \in \mathcal{U}$) learned from attributed-oriented module, i.e., $ \boldsymbol{C} = \{\boldsymbol{C}_1, \boldsymbol{C}_2 \cdots,\boldsymbol{C}_K $ \},  where $\boldsymbol{C}_k$ denotes the $k$-th user cluster. Following the meta-learning paradigm, we construct the meta-training support set $\mathcal{S}^{tr}$ = $\{\mathcal{S}^{tr}_1, \mathcal{S}^{tr}_2 \cdots \mathcal{S}^{tr}_K\}$ based on $\boldsymbol{C}$, where $\mathcal{S}^{tr}_k$ is a partial set of items that have been rated by users $u \in \boldsymbol{C}_k$. 
For new users in cold-start scenarios, we allocate them to their closest cluster center in $\boldsymbol{C}$. These new users form a new cluster set $\boldsymbol{C}'$, and we construct the meta-testing support set $\mathcal{S}^{te}$ according to $\boldsymbol{C}'$. The set $\mathcal{S}^{te}_k$ consists of a subset of items interacted by new users.

Likewise, we can generate the meta-training query set $\mathcal{Q}^{tr}$, which contains user-item pairs simulated as unseen interactions. $\mathcal{Q}^{tr}$ is used to accumulate the task loss in meta-training, while the meta-testing query set $\mathcal{Q}^{te}$ is constructed to evaluate the recommendation results in meta-testing. It should be noted that the support set and query set are mutually exclusive in each task $T_k$ (i.e., $\mathcal{S}_k \bigcap \mathcal{Q}_k = \emptyset$).

\subsection{Augmented Graph Neural Network}


\subsubsection{Graph Embedding Propagation}
After obtaining $K$ different tasks, we construct the user-item bipartite graph and leverage interactions of users in the $k$-th task as training data for $\boldsymbol{T}_k$. We perform GCN to capture the high-order structural information of the interaction graph. Firstly, we randomly initialize free embedding vectors $\boldsymbol{e}_u \in \mathbb{R}^d$ ($\boldsymbol{e}_i \in \mathbb{R}^d$) to denote a user $u$ (item $i$), where $d$ denotes the embedding dimension. Then we perform the light graph convolution (LGC) \cite{LightGCN} from user $u$'s neighbors, which is formulated as follows,
\begin{equation}
\boldsymbol{e}_u^{(l + 1)} = \sum_{i \in \mathcal{N}_u}\frac{1}{\sqrt{|\mathcal{N}_u||\mathcal{N}_i|}} \boldsymbol{e}_i^{(l)}, 
\label{eq:u_agg}
\end{equation} 
where $\boldsymbol{e}_u^{(l+1)}$ ($\boldsymbol{e}_i^{(l+1)}$) denotes the embedding of node $u$ ($i$) in layer $l+1$, memorizing the messages from its $l$-layer neighbors; $\mathcal{N}_u$ ($\mathcal{N}_i$) denotes the set of neighbors of node $u$ ($i$);  $|\mathcal{N}_u|$ ($|\mathcal{N}_i|$) denotes the neighbor size of node $u$ ($i$); $\boldsymbol{e}_u^{(0)}$ = $\boldsymbol{e}_u$ and $\boldsymbol{e}_i^{(0)}$ = $\boldsymbol{e}_i$. Similarly, we can obtain item $i$'s embedding through LGC. The purpose of this module is to learn more effective embedding only through the information propagation on user-item bipartite graph.

\subsubsection{Graph Enhanced Generator}
Considering that the limited interactions of users are insufficient to learn expressive representation in cold-start scenarios, we further adopt two strategies to generate potential interactions for users.

In the first strategy, we mine the structure of the interaction graph and capture potential dependencies to the user-item pair that have not appeared in the graph. Specifically, we expand interactions that are similar to users' existing interacted items based on the graph embedding. From the perspective of Occam’s razor, we adopt a vanilla designed, weighted inner production to measure the interest of the user $u$ to the item $i$ as,
\begin{equation}
\boldsymbol{E}_{u,i}^1 = \sigma( \sum_{j \in \mathcal{N}_u} \boldsymbol{e}_j^{(L)} \boldsymbol{W}_g \boldsymbol{e}_i^{(L)}),
\end{equation} 
where $\boldsymbol{e}_j^{(L)}$ refers to the final $L$-th layer embedding of item $j$, $\boldsymbol{W}_g$ refers to the matrix parameter capturing the structure information between users and items; and $\sigma$ is the sigmoid function. Different from ItemCF \cite{ItemCF}, the strategy not only leverages the neighborhood information but also captures the high-order relationships in the bipartite graph.

In the second strategy, we utilize interacted items to represent users and incorporate potential items according to their attribute similarity with users. We also apply a weighted inner production to define the similarity between user $u$ and item $i$ as, 
\begin{equation}
\boldsymbol{E}_{u,i}^2 = \sigma( \sum_{j \in \mathcal{N}_u} \boldsymbol{z}_j \boldsymbol{W}_a \boldsymbol{z}_i),
\end{equation}
where $\boldsymbol{W}_a$ refers to the matrix parameter capturing the attribute information between users and items; and $\sigma$ is the sigmoid function. We employ a hyper-parameter $\alpha$ in the range of $\left[0,1\right]$ to balance the two strategies,
\begin{equation}
\boldsymbol{E}_{u,i} = \alpha  \boldsymbol{E}_{u,i}^1 + (1 - \alpha) \boldsymbol{E}_{u,i}^2.
\label{gen}
\end{equation}

The loss function for training the potential interaction generator is as follows,
\begin{equation}
\mathcal{L}_{gen} = \left \| \boldsymbol{E} - \boldsymbol{A} \right \|^2.
\label{gen_loss}
\end{equation}
We put the generated edges into the adjacency matrix by setting a threshold $t$,
\begin{equation}
    \boldsymbol{\hat{A}}\left[ u, i\right]=
    \begin{cases}
    1,& \text{if $\boldsymbol{E}_{u,i}$ > $t$;}\\
    0,& \text{otherwise.}
    \end{cases}
\label{eq:new_A}
\end{equation}
where $\boldsymbol{\hat{A}}$ is the generated adjacency matrix. We utilize $\boldsymbol{\hat{A}}$ to augment the interaction, which is beneficial to alleviate the cold-start problem.

\subsubsection{Model Prediction}
In order to model the representation in a more fine-grained way, we concatenate the final $L$-th layer graph embedding of user $u$ with its corresponding attribute embedding as follows, 
\begin{equation}
\boldsymbol{f}_u = \boldsymbol{e}_u^{(L)} \oplus \boldsymbol{z}_u \ ,
\label{eq:final}
\end{equation} 
where $\oplus$ denotes concatenation operation. Analogously, we can obtain the final embedding $\boldsymbol{f}_i$ of item $i$.
We adopt the inner product to estimate the user's preference towards the target item,
\begin{equation}
\hat{y}_{ui} = \boldsymbol{f}_u^T \boldsymbol{f}_i \ .
\label{eq:prediction}
\end{equation} 
We utilize the Bayesian Personalized Ranking (BPR) loss \cite{rendle2009bpr} to optimize model parameters, which is a pairwise loss that encourages the prediction of an observed interaction to be assigned higher than its unobserved ones. The prediction loss is defined as follows,
\begin{equation}
\mathcal{L}_{pre} = \sum_{(u,i,j) \in \mathcal{D}} -{\rm log}\sigma(\hat y_{ui} - \hat y_{uj}),
\label{eq:bpr_loss}
\end{equation} 
where $\mathcal{D}=\{(u,i,j)|(u,i) \in \mathcal{D}^+, (u,j) \in \mathcal{D}^- \}$ denotes the pairwise training data;  $\mathcal{D}^+$ denotes the observed interactions and $\mathcal{D}^-$ denotes the unobserved interactions. $\sigma$ is the sigmoid function.

\subsection{Contrastive Regularization}
Optimizing a ranking-motivated loss \cite{Hao0FX020, bojchevski2018deep} is effective to capture the relationships between each pair of training samples. Contrastive learning \cite{hadsell2006dimensionality} is one kind of ranking-motivated loss. As an extension of Information Maximization (InfoMax) principle \cite{linsker1988self}, contrastive learning learns representations by maximizing the Mutual Information (MI), i.e., contrasting positive pairs with corresponding negative-sampled pairs. We have a similar learning objective in our problem that is to pull the latent attribute embeddings in the same cluster together while pushing the latent attribute embeddings in different clusters far away from each other.
We design a task-wise contrastive regularization based on the attribute embeddings to enhance the latent clustering knowledge. More specifically, we treat attributes in the same task as the positive pairs, denoted by $\{(\boldsymbol{z}_u, \boldsymbol{z}_{u'})|u, u' \in \boldsymbol{C}_k\}$, and attributes in different tasks as negative pairs, denoted by $\{(\boldsymbol{z}_u, \boldsymbol{z}_{v})|u \in \boldsymbol{C}_k, v \notin \boldsymbol{C}_k\}$. The supervision of positive pairs retains consistency in the same task. Meanwhile, the supervision of negative pairs enhances the discriminatory capability of different tasks. Following the contrastive design paradigm \cite{chen2020simple}, we propose the contrastive regularization to maximize MI in task-wise as follows,
\begin{equation}
\begin{aligned}
\mathcal{L}_{MI} =  \sum_{k = 1}^K \sum_{u \in \boldsymbol{C}_k}-{\rm{log}}\frac{{\rm{exp}}({\rm{sim}}(z_u, z_{u'})/\tau)}{\sum_{v \notin \boldsymbol{C}_k}{\rm{exp}}({\rm{sim}}(z_u, z_v)/\tau)}  , 
\label{eq:con_loss}
\end{aligned}
\end{equation} 
where ${\rm{sim}} (\cdot)$ is the function to measure the similarity between two vectors and $\tau$ is the hyper-parameter for softmax temperature. There are many choices of ${\rm{sim}} (\cdot)$, such as cosine similarity, dot product, etc. We observe that using the cosine similarity can achieve good results in our model. The summation in the denominator can be approximated by in-batch negative samples with mini-batch training.

\begin{algorithm}[t]
	\caption{Meta-training of TMAG}   
	\label{training}     
	\begin{algorithmic}[1] 
	\Require  the graph $\mathcal{G}$; learning rates $\alpha$ and $\beta$;
    \State Randomly initialize the two attribute-oriented autoencoders, graph enhanced generator and other parameters (e.g., $\boldsymbol{\theta}$)
    \State Fix other parts, train the attribute-oriented autoencoders until convergence by Equation \eqref{auto_loss} and obtain $\boldsymbol{z}_u$, $\boldsymbol{z}_i$
    \State Construct aligned training tasks $\mathcal{T}^{tr}$, each task $T_{k} \in \mathcal{T}^{tr}$ consisting of a support set $\mathcal{S}_k^{tr}$ and a query set $\mathcal{Q}_k^{tr}$
    \While{not convergence}
    
    \State Randomly select a task $T_k \in \mathcal{T}^{tr}$
    \State Compute $\boldsymbol{e}_u^{(L)}$, $\boldsymbol{e}_i^{(L)}$ with $\mathcal{G} $ by Equation \eqref{eq:u_agg} 
    \State Calculate the final embedding $\boldsymbol{f}_u$, $\boldsymbol{f}_i$ by Equation \eqref{eq:final}
    \State Evaluate $\mathcal{L}_{T_k}(\boldsymbol{\theta}_k, \boldsymbol{S}_k^{tr})$ by Equation \eqref{eq:total_loss}
    \State Local update $\boldsymbol{\theta}_k' = \boldsymbol{\theta}_k - \alpha \nabla_{\boldsymbol{\theta}}\mathcal{L}_{T_k}(\boldsymbol{\theta}_k, \boldsymbol{S}_k^{tr})$
    \State Evaluate $\mathcal{L}_{T_k}(\boldsymbol{\theta}_k', \boldsymbol{Q}_k^{tr})$ with query set $\boldsymbol{Q}_k^{tr}$
    \State Global update $\boldsymbol{\theta} = \boldsymbol{\theta} - \beta \nabla_{\boldsymbol{\theta}}\mathcal{L}_{T_k}(\boldsymbol{\theta}_k', \boldsymbol{Q}_k^{tr})$
    \State Generate $\boldsymbol{\hat{A}}$ by Equation \eqref{eq:new_A} and update $\mathcal{G}$ 
    \EndWhile
   \end{algorithmic} 
\end{algorithm}

\subsection{Optimization}
To improve recommendation in cold-start scenarios, we leverage a multi-task training strategy to jointly optimize the main recommendation task (cf. Equation \eqref{eq:bpr_loss}), the interaction generator task (cf. Equation \eqref{gen_loss}) and the self-supervised learning task (cf. Equation \eqref{eq:con_loss}).
\begin{equation}
\mathcal{L} = \mathcal{L}_{pre} + \lambda_1 \mathcal{L}_{gen} + \lambda_2 \mathcal{L}_{MI} + \lambda_3 || \boldsymbol{\Theta} ||^2,
\label{eq:total_loss}
\end{equation} 
where $\boldsymbol{\Theta}$ denotes all trainable parameters in $\mathcal{L}_{pre}$ and $\mathcal{L}_{gen}$ since $\mathcal{L}_{MI}$ adds no additional parameters; $\lambda_1$, $\lambda_2$ and $\lambda_3$ parameterize the weights of different losses.

Following the MAML framework, we perform a few second-order gradient descent updates in the meta-training process to obtain suitable initial parameter $\boldsymbol{\theta}$ for users and items. Then, the representation of new users and items can be rapidly adapted with only a few interactions. The whole algorithm is summarized in Algorithm \ref{training}. Taking $k$-th task $T_k$ as an example, we perform a few gradient descent updates according to the loss function defined in Equation \eqref{eq:total_loss}, i.e., inner-loop update. For simplicity, we perform one update on $\boldsymbol{S}_k^{tr}$ as follows:
\begin{equation}
\boldsymbol{\theta}_k' = \boldsymbol{\theta}_k - \alpha \nabla_{\boldsymbol{\theta}}\mathcal{L}_{T_k}(\boldsymbol{\theta}_k, \boldsymbol{S}_k^{tr}),
\label{eq:inner}
\end{equation} 
where $\mathcal{L}_{T_k}$ denotes the loss function of $T_k$; $\alpha$ is the inner-loop learning rate and $\boldsymbol{\theta}_k'$ denotes the new parameter. 

In the outer-loop process, we calculate the query set loss on $\boldsymbol{Q}_k^{tr}$ to update the initial parameter $\boldsymbol{\theta}$ as:
\begin{equation}
\boldsymbol{\theta} = \boldsymbol{\theta} - \beta \nabla_{\boldsymbol{\theta}}\mathcal{L}_{T_k}(\boldsymbol{\theta}_k', \boldsymbol{Q}_k^{tr}),
\label{eq:outer}
\end{equation} 
where $\beta$ is the outer-loop learning rate.

\section{Experiments}
In this section, we conduct extensive experiments to answer the following four questions:

\begin{itemize}[leftmargin=*]
\item \textbf{RQ1}: How does TMAG perform compared to state-of-the-art cold-start recommendation methods?
\item \textbf{RQ2}: How do different components of TMAG affect the recommendation performance?
\item \textbf{RQ3}: How is TMAG impacted by the sparsity of the support set and its hyper-parameters?
\item \textbf{RQ4}: Can TMAG provide qualitative analyses of learned representations with regard to aligned tasks?
\end{itemize}

\subsection{Experimental Setup}

\subsubsection{Experimental Setup}
We conduct extensive experiments on the following three datasets, which are provided by \cite{MetaHIN}. 

\begin{itemize}[leftmargin=*]
\item DBook\footnote{https://book.douban.com}: This is a widely used dataset for book recommendation obtained from Douban. We divide books into existing and new items based on their publishing year, with a roughly 8:2 ratio. Due to the lack of temporal information about users, we randomly picked 80\% of users as existing users and the remaining 20\% as new users.

\item MovieLens\footnote{https://grouplens.org/datasets/movielens/}: This is a widely-used benchmark dataset published by GroupLens for movie recommendation, where movies ratings were released from 1919 to 2000. We divide movies into old items (released before 1998) and new items (issued between 1998 and 2000) with a roughly 8:2 ratio. To designate new users in the MovieLens dataset, we arrange users by their first rating timestamp, with the most recent 20\% of users considered to be new to MovieLens.

\item Yelp\footnote{https://www.yelp.com/dataset/challenge}: This dataset is from Yelp business and is widely used for recommendation. Users who joined Yelp before May 1, 2014 are considered as existing users, while the rest are considered as new users. Similarly, we divide firms as old or new based on the date they were first rated. The ratio of new users to existing users is approximately 8:2.
\end{itemize}

In a nutshell, for each dataset, we split users and items into two groups: existing and new, based on the first user interacting time and first item releasing time. Then we separate each dataset into meta-training and meta-testing data. We consider existing user feedback on existing items as meta-training data and the rest as meta-testing data. We randomly select 10\% of them as the validation set. The meta-testing data is divided into three tasks: (1)Task1: Recommend existing items for new users; (2)Task2: Recommend new items for existing users; (3)Task3: Recommend new items for new users. We only add the additional recall method in cold-start scenarios to learn better initial embeddings for new users and items. Therefore, the recommendation results of warm users and items are not influenced, and we do not report the performance.

On these three datasets, users' ratings on items are explicit and the ratings range from 1-5. Since we evaluate on implicit feedback, following the previous work \cite{kang2018self, sachdeva2019sequential}, we regard the ratings that are higher than 3 as positive feedback. Their statistics are summarized in Table \ref{tab:data}. Following the previous work \cite{MeLU, MetaHIN, MvDGAE}, we filter users whose interactions are more than 100 or less than 13. For each user $u$, we randomly select 10 interacted items as the query set, and the rest items are used as the support set. We will study how TMAG is impacted by the size of the support set in Section \ref{sec:sparsity}.

\begin{table}[t]
\small
\setlength{\abovecaptionskip}{0.1cm}
\setlength{\belowcaptionskip}{-0cm}
\caption{Statistics of Preprocessed Datasets}
\label{tab:data}
\begin{tabular}{c|c|c|c}
\hline  & DBook        & MovieLens   & Yelp         \\ \hline \hline
Users         &10,592           &6,040       & 51,624   \\
Items         &20,934          &3,881       & 34,199   \\
Interactions  & 649,381     & 1,000,209    & 1,301,869     \\
Sparsity     & 99.71\%   & 95.73\%  & 92.63\%      \\ \hline 
User contents & \begin{tabular}[c]{@{}c@{}}Group,\\ Location\end{tabular}  & \begin{tabular}[c]{@{}c@{}}Age, Occupation,\\ Gender, Zip code\end{tabular}         & \begin{tabular}[c]{@{}c@{}}Fans,\\  Friends\end{tabular}                     \\ \hline
Item contents  & \begin{tabular}[c]{@{}c@{}}Publisher,\\ Author, Year\end{tabular} & \begin{tabular}[c]{@{}c@{}}Actor, Year,\\ Director, Genre\end{tabular}     & \begin{tabular}[c]{@{}c@{}}Category, City, \\ Postal code, State\end{tabular} \\ \hline
\end{tabular}
\end{table}

\subsubsection{Baselines}
We compare our model against three different types of baselines, namely, traditional methods (MF and NeuMF), GNN-based methods (NGCF, GraphSAINT and LightGCN), and cold-start methods (MeLU, MetaHIN and CLCRec). For traditional methods and GNN-based methods, we train the base model on meta-training data. We further fine-tune the base model using support sets from the meta-testing data to adapt to the cold-start scenarios.

\begin{itemize}[leftmargin=*]
\item MF \cite{MF}: It utilizes conventional matrix factorization optimized by BPR loss for item recommendation, which only leverages direct interactions between users and items as the target value of the interaction function.
\item NeuMF \cite{NCF}: It is the most representative deep neural network based method for collaborative filtering, which unifies MLP and matrix factorization to a general framework to learn the embedding of users and items.
\item NGCF \cite{NGCF}: It is a highly effective collaborative filtering method based on GCN that captures collaborative signal by propagating graph embeddings. It injects the collaborative signal into the embedding process in an explicit manner.
\item GraphSAINT \cite{zeng2019graphsaint}: It is a general GNN model, which builds a complete GCN from sampled subgraph. And it proposes normalization technique to eliminate bias, and sampling algorithms for variance reduction.
\item LightGCN \cite{LightGCN}: It improves NGCF by discarding the feature transformation and nonlinear activation, which learns user and item embeddings by linearly propagating them on the user-item interaction graph.
\item MeLU \cite{MeLU}: It applies MAML with a few steps of gradient updates to alleviate the user cold-start problem. It only considers user-item interactions and features.
\item MetaHIN \cite{MetaHIN}: It is a meta-learning method that leverages the rich semantic of HIN to capture richer semantics via higher-order graph structures.
\item CLCRec \cite{wei2021contrastive}: It maximizes the mutual information between collaborative embeddings of users and items. Besides, it maximizes the mutual information between collaborative embeddings and feature representations of items. 
\end{itemize}

\begin{table*}[t]
\small
\setlength{\abovecaptionskip}{0.2cm}
\setlength{\belowcaptionskip}{-0cm}
\caption{Comparison between our proposed model and other baselines at top-10. R is short for \emph{Recall}, ND for \emph{NDCG}, M for \emph{MAP}. Boldface denotes the highest score and underlines denotes the best performing baselines.}
\label{data}
\centering

\begin{tabular}{c||c||ccc||ccc||ccc}
\hline
\multirow{2}{*}{Scenario}                                                                            & \multirow{2}{*}{Model} & \multicolumn{3}{c||}{DBook}                          & \multicolumn{3}{c||}{MovieLens}                      & \multicolumn{3}{c}{Yelp}                            \\ \cline{3-11} 
                                                                                                     &                        & R@10            & ND@10           & M@10            & R@10            & ND@10           & M@10            & R@10            & ND@10           & M@10            \\ \hline
\multirow{10}{*}{\begin{tabular}[c]{@{}c@{}}Task1\\ (Existing items \\ for new users)\end{tabular}} & MF                     & 0.0834          & 0.0941          & 0.0388          & 0.1403          & 0.1588          & 0.0718          & 0.0517          & 0.0552          & 0.0193          \\
                                                                                                     & NeuMF                  & 0.0852          & 0.0955          & 0.0395          & 0.1439          & 0.1612          & 0.0729          & 0.0520          & 0.0557          & 0.0196          \\ \cline{2-11} 
                                                                                                     & NGCF                   & 0.0862          & 0.0980          & 0.0417          & 0.1524          & 0.1729          & 0.0797          & 0.0544          & 0.0607          & 0.0223          \\
                                                                                                     & GraphSAINT             & 0.0998          & 0.1149          & 0.0497          & 0.1980          & 0.2233          & 0.1066          & {\ul 0.0634}    & {\ul 0.0706}    & {\ul 0.0268}    \\
                                                                                                     & LightGCN               & {\ul 0.1007}    & {\ul 0.1175}    & {\ul 0.0516}    & 0.2022          & 0.2279          & 0.1106          & 0.0630          & 0.0703          & 0.0266          \\ \cline{2-11} 
                                                                                                     & MeLU                   & 0.0854          & 0.0969          & 0.0398          & 0.1452          & 0.1634          & 0.0742          & 0.0530          & 0.0564          & 0.0203          \\
                                                                                                     & MetaHIN                & 0.0860          & 0.0973          & 0.0399          & 0.1503          & 0.1685          & 0.0768          & 0.0536          & 0.0579          & 0.0211          \\
                                                                                                     & CLCRec                 & 0.0998          & 0.1106          & 0.0465          & {\ul 0.2077}    & {\ul 0.2387}    & {\ul 0.1163}    & 0.0551          & 0.0559          & 0.0192          \\ \cline{2-11} 
                                                                                                     & $\textbf{TMAG}$                   & $\boldsymbol{0.1046}^*$  & $\boldsymbol{0.1212}^*$ & $\boldsymbol{0.0530}^*$ & $\boldsymbol{0.2115}^*$ & $\boldsymbol{0.2451}^*$ & $\boldsymbol{0.1210}^*$ & $\boldsymbol{0.0665}^*$ & $\boldsymbol{0.0732}^*$ & $\boldsymbol{0.0276}^*$ \\ \cline{2-11} 
                                                                                                     & \%Improv.              & 3.87\%          & 3.15\%          & 2.71\%          & 1.83\%          & 2.68\%          & 4.04\%          & 4.89\%          & 3.68\%          & 2.99\%          \\ \hline \hline
\multirow{10}{*}{\begin{tabular}[c]{@{}c@{}}Task2\\ (New items for \\ existing users)\end{tabular}}  & MF                     & 0.1278          & 0.1450          & 0.0667          & 0.2336          & 0.2706          & 0.1359          & 0.0554          & 0.0582          & 0.0214          \\
                                                                                                     & NeuMF                  & 0.1292          & 0.1470          & 0.0680          & 0.2362          & 0.2734          & 0.1377          & 0.0562          & 0.0576          & 0.0206          \\ \cline{2-11} 
                                                                                                     & NGCF                   & 0.1499          & 0.1691          & 0.0784          & 0.2666          & 0.3086          & 0.1630          & 0.0745          & 0.0816          & 0.0316          \\
                                                                                                     & GraphSAINT             & 0.1664          & 0.1861          & 0.0890          & 0.2729          & 0.3163          & 0.1682          & 0.0845          & {\ul 0.0938}    & {\ul 0.0382}    \\
                                                                                                     & LightGCN               & {\ul 0.1673}    & {\ul 0.1868}    & {\ul 0.0896}    & {\ul 0.2768}    & {\ul 0.3198}    & {\ul 0.1715}    & {\ul 0.0849}    & 0.0933          & 0.0378          \\ \cline{2-11} 
                                                                                                     & MeLU                   & 0.1370          & 0.1526          & 0.0705          & 0.2466          & 0.2891          & 0.1492          & 0.0584          & 0.0657          & 0.0256          \\
                                                                                                     & MetaHIN                & 0.1392          & 0.1549          & 0.0717          & 0.2509          & 0.2930          & 0.1518          & 0.0602          & 0.0677          & 0.0263          \\
                                                                                                     & CLCRec                 & 0.1437          & 0.1620          & 0.0741          & 0.2417          & 0.2840          & 0.1455          & 0.0809          & 0.0844          & 0.0307          \\ \cline{2-11} 
                                                                                                     & $\textbf{TMAG}$                   & $\boldsymbol{0.1732}^*$ & $\boldsymbol{0.1969}^*$ & $\boldsymbol{0.0968}^*$ & $\boldsymbol{0.2871}^*$ & $\boldsymbol{0.3314}^*$ & $\boldsymbol{0.1791}^*$ & $\boldsymbol{0.0906}^*$ & $\boldsymbol{0.0998}^*$ & $\boldsymbol{0.0397}^*$ \\ \cline{2-11} 
                                                                                                     & \%Improv.              & 3.53\%          & 5.41\%          & 8.04\%          & 3.72\%          & 3.63\%          & 4.43\%          & 6.71\%          & 6.40\%          & 3.93\%          \\ \hline \hline
\multirow{10}{*}{\begin{tabular}[c]{@{}c@{}}Task3\\ (New items\\ for new users)\end{tabular}}       & MF                     & 0.0983          & 0.1085          & 0.0457          & 0.2256          & 0.2649          & 0.1319          & 0.0280          & 0.0308          & 0.0109          \\
                                                                                                     & NeuMF                  & 0.1026          & 0.1128          & 0.0471          & 0.2281          & 0.2650          & 0.1316          & 0.0288          & 0.0311          & 0.0109          \\ \cline{2-11} 
                                                                                                     & NGCF                   & 0.1152          & 0.1252          & 0.0552          & 0.2366          & 0.2755          & 0.1388          & 0.0667          & 0.0687          & 0.0264          \\
                                                                                                     & GraphSAINT             & 0.1248          & 0.1340          & 0.0571          & 0.2537          & 0.2951          & 0.1519          & {\ul 0.0698}    & {\ul 0.0717}    & {\ul 0.0266}    \\
                                                                                                     & LightGCN               & {\ul 0.1263}    & {\ul 0.1363}    & {\ul 0.0586}    & {\ul 0.2568}    & {\ul 0.2971}    & {\ul 0.1536}    & 0.0686          & 0.0704          & 0.0262          \\ \cline{2-11} 
                                                                                                     & MeLU                   & 0.1056          & 0.1187          & 0.0497          & 0.2326          & 0.2700          & 0.1354          & 0.0328          & 0.0329          & 0.0113          \\
                                                                                                     & MetaHIN                & 0.1077          & 0.1200          & 0.0503          & 0.2346          & 0.2726          & 0.1369          & 0.0339          & 0.0337          & 0.0115          \\
                                                                                                     & CLCRec                 & 0.1068          & 0.1195          & 0.0506          & 0.2162          & 0.2525          & 0.1244          & 0.0607          & 0.0627          & 0.0223          \\ \cline{2-11} 
                                                                                                     & $\textbf{TMAG}$                   & $\boldsymbol{0.1325}^*$ & $\boldsymbol{0.1427}^*$ & $\boldsymbol{0.0619}^*$ & $\boldsymbol{0.2649}^*$ & $\boldsymbol{0.3062}^*$ & $\boldsymbol{0.1612}^*$ & $\boldsymbol{0.0723}^*$ & $\boldsymbol{0.0749}^*$ & $\boldsymbol{0.0281}^*$ \\ \cline{2-11} 
                                                                                                     & \%Improv.              & 4.91\%          & 4.70\%          & 5.63\%          & 3.15\%          & 3.06\%          & 4.95\%          & 3.58\%          & 4.46\%          & 5.64\%          \\ \hline
\end{tabular}

\begin{tablenotes}
   \footnotesize
     \item \hspace*{1cm} * indicates the improvements of TMAG over the best baseline are statistically significant (i.e., one-sample t-test with \emph{p} < 0.05) .
 \end{tablenotes}

\end{table*}

\subsubsection{Evaluation Metrics}
Cold-start recommendation can be considered as making top-n recommendations for users. We select three widely-used ranking metrics to evaluate the performance:Recall@$K$, Normalized Discounted Cumulative Gain (NDCG)@$K$ and Mean Average Precision (MAP)@$K$:

\begin{itemize}[leftmargin=*]
    \item \emph{Recall}: It means the proportion of actually interacted items that appear within the top-$K$ ranking list.
    \item \emph{NDCG}: It is a standard ranking metric that reflects both the correlation and position for each recommended item. 
    \item \emph{MAP}: The mean of the average precision scores for each user. And precision is the proportion of items actually interacted in the recommended list.
\end{itemize}

Here we use $K$ = 10. The two-tailed unpaired t-test~\cite{bhattacharya2002median} is performed to detect significant differences between TMAG and the best baseline.

\subsubsection{Implementation Details}
For MF and NeuMF, we use the codes in the NeuRec\footnote{https://github.com/wubinzzu/NeuRec} library. The source codes of NGCF, GraphSAINT, LightGCN, MeLU, MetaHIN and CLCRec have been released. And all of the settings are following the original suggested settings. We change parts of input data and evaluations to fit our experiment setting. We implement our TMAG model in Tensorflow \footnote{https://www.tensorflow.org/}. To make comparisons fair, the embedding size of users and items is fixed to 64 for all models. We train all the models with Adam optimizer \cite{Adam}, and we use the Xavier initializer \cite{glorot2010understanding} to initialize model parameters. We also apply the early stop strategy to prevent over-fitting. In terms of hyper-parameters, we apply a grid search for hyper-parameters: the inner-loop and outer-loop learning rate is tuned amongst $\{0.0001, 0.0005, 0.001, 0.005\}$; the cluster number ranges from 10 to 50 with the step length 10; the number of inner updates vary from 1 to 5; we search the temperature coefficient in range of $\{0.1, 0.2, 0.5, 1.0\}$; the augmentation threshold is tuned amongst $\{0.5, 0.6, 0.7, 0.8, 0.9\}$; $\lambda_1$, $\lambda_2$ and $\lambda_3$ are searched in $\{0.005, 0.01, 0.05, 0.1, 0.5, 1.0\}$.

\subsection{Performance Comparison (RQ1)}

Table \ref{data} shows the top-10 recommendation performance on three datasets. From the table, we have the following observes: 
\begin{itemize}[leftmargin=*]
\item In the first type of baseline methods, we can find that the performance of MF method is slightly lower than the deep learning method NeuMF, highlighting the critical role of nonlinear feature interactions between user and item embeddings. However, neither MF nor NeuMF models the generalization ability to new users and items, which leads to poor performance in cold-start scenarios. 

\item In the second type of baseline methods, GNN-based methods achieve great improvements in all datasets and scenarios over traditional methods, especially LightGCN and GraphSAINT. Because we focus on the incomplete cold-start problem, GNN-based model is capable of improving cold-start user and item representations through exploring the high-order relationships. And on Yelp dataset, we can observe that GraphSAINT beats LightGCN slightly. The possible reason is that the limited data on Yelp causes over-fitting and thus it restricts the performance of the model, while GraphSAINT alleviates the problem by utilizing subgraphs to promote robustness.
    
\item In the third type of baseline methods, CLCRec achieves competitive performance because it captures more information related to the collaborative signal by maximizing the mutual information. As for meta-learning-based methods, they show superior performance than traditional methods owing to the well-designed training process that results in a personalized parameter initialization for new users. MetaHIN surpasses MeLU in every scenario because it incorporates multiple semantics obtained from higher-order structures such as meta-paths. Since these methods ignore the high-order graph structure, the performance is inferior to those GNN-based methods.

\item TMAG consistently outperforms all baseline methods in all scenarios across the datasets. For instance, TMAG improves over the best baseline w.r.t. Recall@10 by 3.53-4.91\%, 1.83-3.72\%, and 3.58-6.71\% on three datasets. The reasons are concluded as follows: 1) The aligned task extracts latent clustering knowledge among similar users that can be quickly adapted to new users.  2) The graph neural network captures the high-order user-item relationships, while MeLU and MetaHIN ignore the graph structure information in user-item graph. And the interaction augmentation mitigates the impact of sparse interactions. 3) The task-wise contrastive regularization enhances the latent clustering prior knowledge. We introduce in detail in the ablation study.
\end{itemize}

\begin{table}[]
\small
    \setlength{\abovecaptionskip}{0.1cm}
    \setlength{\belowcaptionskip}{-0.1cm}
    \caption{Effect of the number of aligned tasks.}
    \label{cluster}
\setlength{\tabcolsep}{1.3mm}{
\begin{tabular}{c|cc|cc|cc}
\hline
\multirow{2}{*}{Task1} & \multicolumn{2}{c|}{DBook}        & \multicolumn{2}{c|}{MovieLens}    & \multicolumn{2}{c}{Yelp}          \\
                       & R@10            & ND@10           & R@10            & ND@10           & R@10            & ND@10           \\ \hline \hline
TMAG-1                 & 0.1015          & 0.1178          & 0.2082          & 0.2392          & 0.0636          & 0.0714          \\
TMAG-10                & 0.1022          & 0.1185          & 0.2094          & 0.2413          & 0.0648          & 0.0721          \\
TMAG-20                & 0.1038          & 0.1193          & 0.2106          & 0.2431          & 0.0662          & 0.0727          \\
TMAG-30                & 0.1039          & 0.1202          & 0.2113          & 0.2443          & \textbf{0.0665} & \textbf{0.0732} \\
TMAG-40                & \textbf{0.1046} & \textbf{0.1212} & \textbf{0.2115} & \textbf{0.2451} & 0.0664          & 0.0731          \\
TMAG-50                & 0.1041          & 0.1207          & 0.2111          & 0.2439          & 0.0660          & 0.0729          \\ \hline
\end{tabular}}
\end{table}

\begin{table}[t]
\small
    \setlength{\abovecaptionskip}{0.1cm}
    \setlength{\belowcaptionskip}{-0.1cm}
    \caption{Effect of the autoencoder.}
    \label{au}
\setlength{\tabcolsep}{1.3mm}{
\begin{tabular}{c|cc|cc|cc}
\hline 
\multirow{2}{*}{Task1} & \multicolumn{2}{c|}{DBook}        & \multicolumn{2}{c|}{MovieLens}    & \multicolumn{2}{c}{Yelp}          \\
                       & R@10            & ND@10           & R@10            & ND@10           & R@10            & ND@10           \\ \hline \hline
w/o AE                 & 0.1030          & 0.1186          & 0.2101          & 0.2443          & 0.0653          & 0.0727          \\
w/ AE                  & \textbf{0.1046} & \textbf{0.1212} & \textbf{0.2115} & \textbf{0.2451} & \textbf{0.0665} & \textbf{0.0732} \\ \hline
\end{tabular}}
\end{table}

\subsection{Ablation Study (RQ2)}
We perform an ablation study to investigate the effectiveness of the proposed TMAG under different circumstances. In the following experiments, we use task1 as the default task and report its performance. The evaluation metrics are R@10 and ND@10.

\subsubsection{Effect of Task Alignment}
To investigate whether TMAG can benefit from task alignment, we search the number of clusters in the range of $\{1, 10, 20, 30, 40, 50\}$. The experimental results are summarized in Table \ref{cluster}, where TMAG-1 denotes that we do not align tasks and regard each user as a single task. Jointly analyzing Table \ref{data} and Figure \ref{cluster}, we have the following observations:

\begin{itemize}[leftmargin=*]
\item Increasing the number of clusters enhances the recommendation results. Obviously, TMAG with task alignment is consistently superior to TMAG-1 in all cases. We attribute the improvement to the fine-grained modeling of similar user groups. 

\item TMAG-40 achieves the best result on MovieLens and DBook and TMAG-30 performs the best on Yelp. The possible reason is that user side information on Yelp is not as diverse as the other two datasets, so it does not need too many clusters to extract the attribute knowledge.

\item TMAG consistently outperforms other approaches when the number of clusters is varied across three datasets. It demonstrates the efficiency of TMAG by extracting latent clustering knowledge and adapting the globally shared meta-knowledge to the latent clustering knowledge that can be rapidly adapted to new users.

\end{itemize}

\begin{table}[t]
\small
    \setlength{\abovecaptionskip}{0.1cm}
    \setlength{\belowcaptionskip}{-0.1cm}
    \caption{Effect of the augmentation variants.}
    \label{aug}
\setlength{\tabcolsep}{1.3mm}{
\begin{tabular}{c|cc|cc|cc}
\hline
\multirow{2}{*}{Task1} & \multicolumn{2}{c|}{DBook}        & \multicolumn{2}{c|}{MovieLens}    & \multicolumn{2}{c}{Yelp}          \\
                       & R@10            & ND@10           & R@10            & ND@10           & R@10            & ND@10           \\ \hline \hline
TMAG-ga                & 0.1034          & 0.1198          & 0.2101          & 0.2438          & 0.0648          & 0.0721          \\
TMAG-a                 & 0.1044          & 0.1209          & 0.2113          & 0.2449          & 0.0661          & 0.0728          \\
TMAG-g                 & 0.1035          & 0.1201          & 0.2103          & 0.2441          & 0.0650          & 0.0725          \\
TMAG                   & \textbf{0.1046} & \textbf{0.1212} & \textbf{0.2115} & \textbf{0.2451} & \textbf{0.0665} & \textbf{0.0732} \\ \hline
\end{tabular}}
 \begin{tablenotes}
           \footnotesize
             \item[$^*$]Subscript notation: TMAG-ga removes the graph and attribute augmentation, TMAG-a removes the attribute augmentation, and TMAG-g removes the graph augmentation.
         \end{tablenotes}
    \vspace{-1em}

\end{table}

\begin{table}[t]
\small
    \setlength{\abovecaptionskip}{0.1cm}
    \setlength{\belowcaptionskip}{-0.1cm}
    \caption{Effect of the contrastive augmentation.}
    \label{con}
\setlength{\tabcolsep}{1.3mm}{
\begin{tabular}{c|cc|cc|cc}
\hline
\multirow{2}{*}{Task1} & \multicolumn{2}{c|}{DBook}        & \multicolumn{2}{c|}{MovieLens}    & \multicolumn{2}{c}{Yelp}          \\
                       & R@10            & ND@10           & R@10            & ND@10           & R@10            & ND@10           \\ \hline \hline
w/o Con                & 0.1034          & 0.1196          & 0.2104          & 0.2439          & 0.0652          & 0.0720          \\
w/ Con                 & \textbf{0.1046} & \textbf{0.1212} & \textbf{0.2115} & \textbf{0.2451} & \textbf{0.0665} & \textbf{0.0732} \\ \hline
\end{tabular}}
\end{table}

\subsubsection{Effect of Autoencoder}
We attempt to understand how the autoencoder facilitates the clustering for aligned tasks. We consider two methods of clustering: 1) directly using the side information of users and items by one-hot encoding; 2) employing an autoencoder to obtain the attribute embeddings. Table \ref{au} describes the results. We can observe that TMAG with autoencoder performs better. We attribute the improvement to the effective learning of attribute representations by the autoencoder.

\subsubsection{Effect of Interaction Augmentation}
In TMAG, we combine the graph structure information and attribute information to augment the adjacency matrix of the graph to alleviate the problem of lacking interactions. To investigate its rationality, we test different settings by removing one type of augmentation or both.

Table \ref{aug} shows the results. We can observe that the best setting is using both two types of 
augmentation. Removing either type will drop the performance. And the graph augmentation consistently outperforms the attribute augmentation, which demonstrates that collaborative signal captured from the graph structure is more beneficial than that from attributes to mine potential interactions. Both two augmentation methods are superior to the variant without augmentation, indicating that the interaction generator helps to learn more expressive representations.

\subsubsection{Effect of Contrastive Regularization}
To study the effectiveness of contrastive regularization, we test our model in two cases: 1) TMAG without contrastive regularization; 2) TMAG with contrastive regularization. Table \ref{con} shows the results. We can observe that TMAG with the contrastive regularization performs better, which indicates the task-wise attribute contrastive learning enhances the latent clustering prior knowledge and enables the model to learn more informative attribute representations.

\begin{figure}[t]
    \setlength{\abovecaptionskip}{-0.1cm}
    \setlength{\belowcaptionskip}{-0.3cm}
    \centering
    \centering
    \subfigcapskip=-6pt
    \subfigure[DBook]{
    \begin{minipage}[c]{0.15\textwidth}
    \centering
	\includegraphics[width=\linewidth]{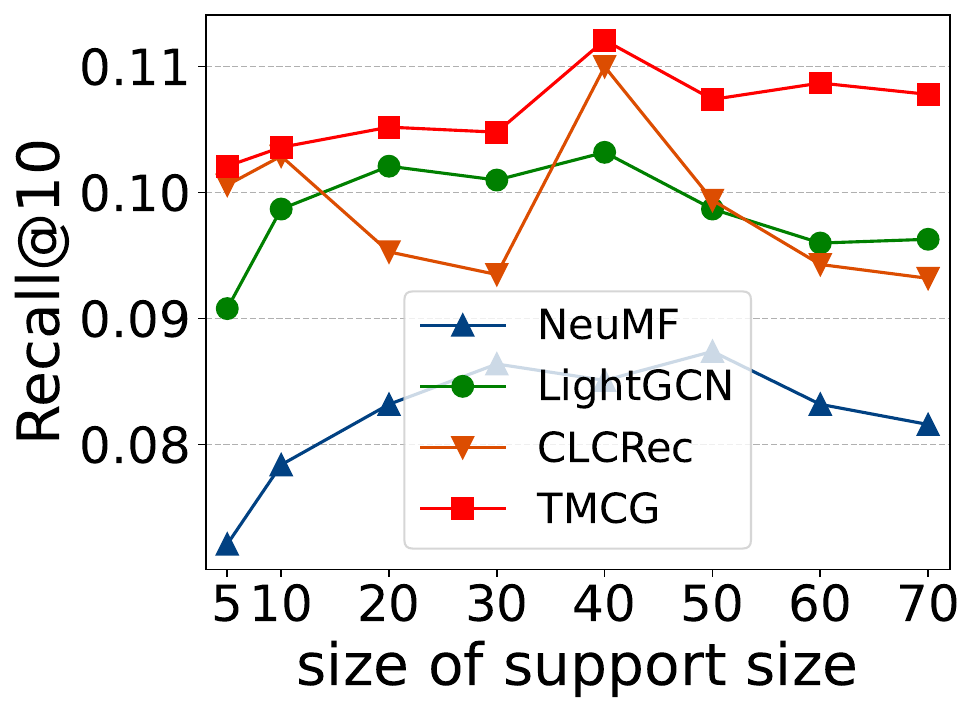}
    \label{fig:ml1m-loss}
    \end{minipage}%
    }
    \subfigcapskip=-6pt
    \subfigure[MovieLens]{
    \begin{minipage}[c]{0.15\textwidth}
    \centering
    \includegraphics[width=\linewidth]{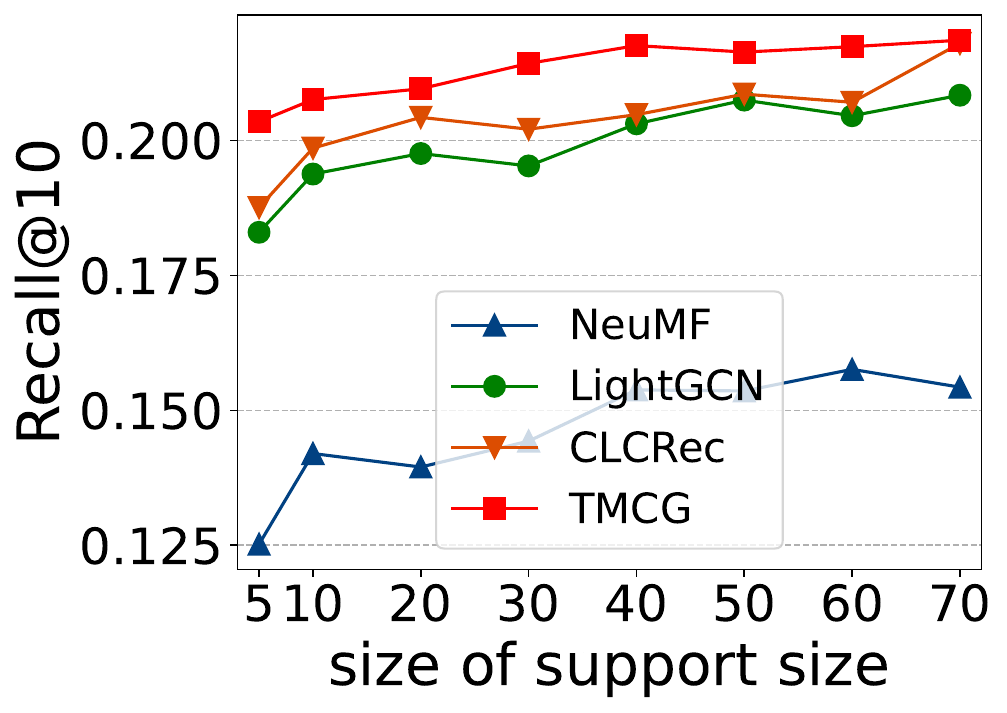}
    \label{fig:yelp-loss}
    \end{minipage}%
    }
    \subfigcapskip=-6pt
    \subfigure[Yelp]{
    \begin{minipage}[c]{0.15\textwidth}
    \centering
    \includegraphics[width=\linewidth]{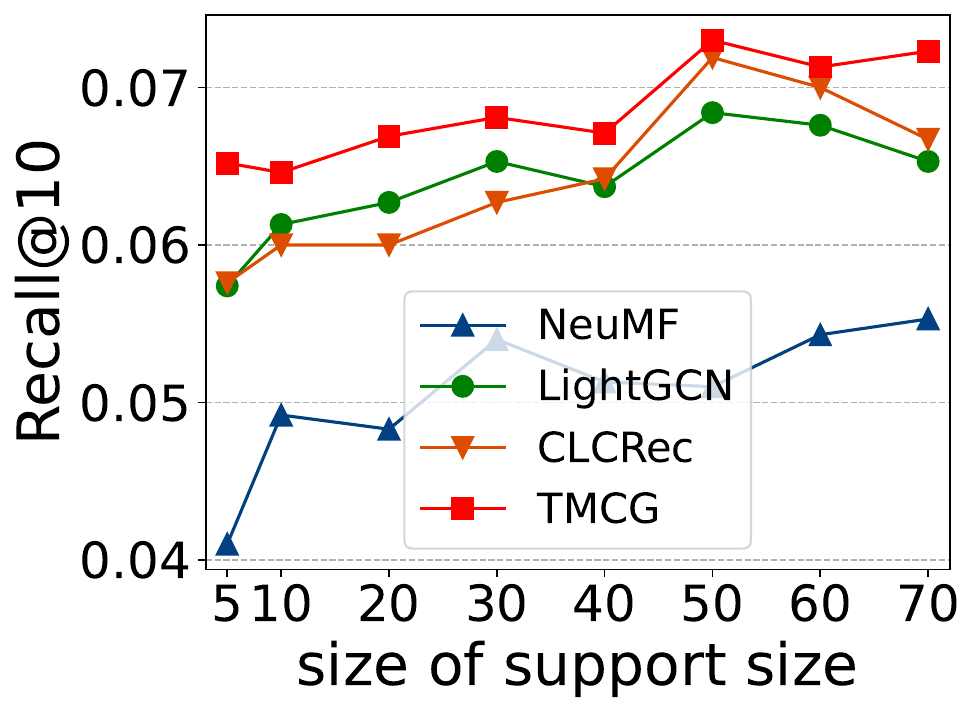}
    \label{fig:amazon-loss}
    \end{minipage}%
    }
    \caption{Impacts of the size of support sets.}
    \label{fig:sparsity}
    
\end{figure}

\begin{figure}[t]
\setlength{\abovecaptionskip}{-0.1cm}
\setlength{\belowcaptionskip}{-0.3cm}
    \centering
    \centering
    \subfigcapskip=-6.5pt
    \subfigure[DBook]{
    \begin{minipage}[c]{0.15\textwidth}
    \centering
    \includegraphics[width=\linewidth]{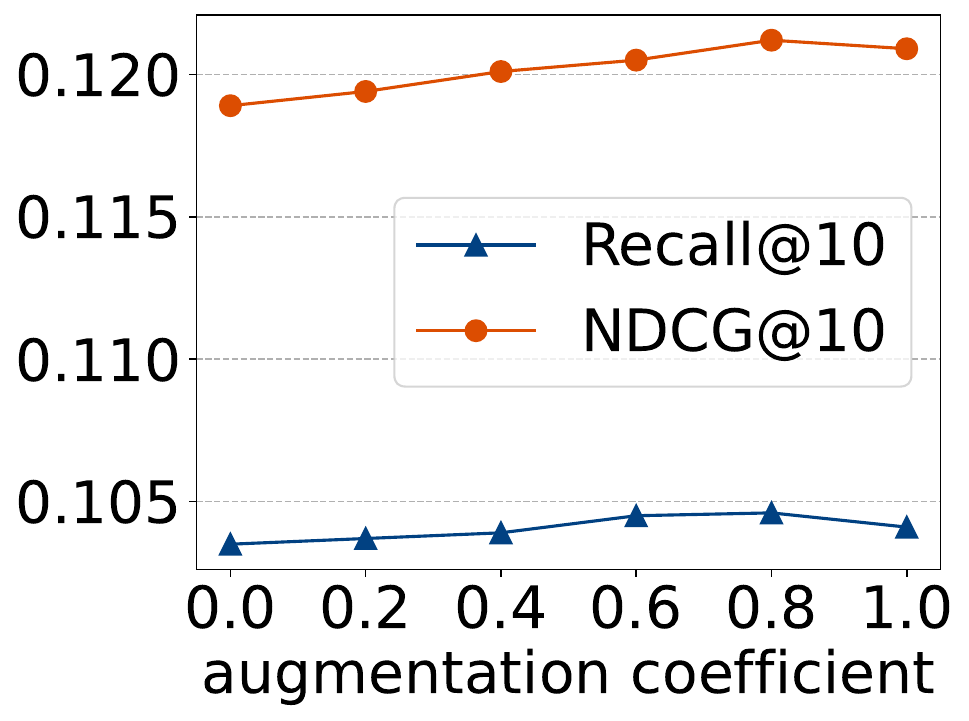}
    \label{fig:ml1m-loss}
    \end{minipage}%
    }
    \subfigcapskip=-8pt
    \subfigure[MovieLens]{
    \begin{minipage}[c]{0.15\textwidth}
    \centering
    \includegraphics[width=\linewidth]{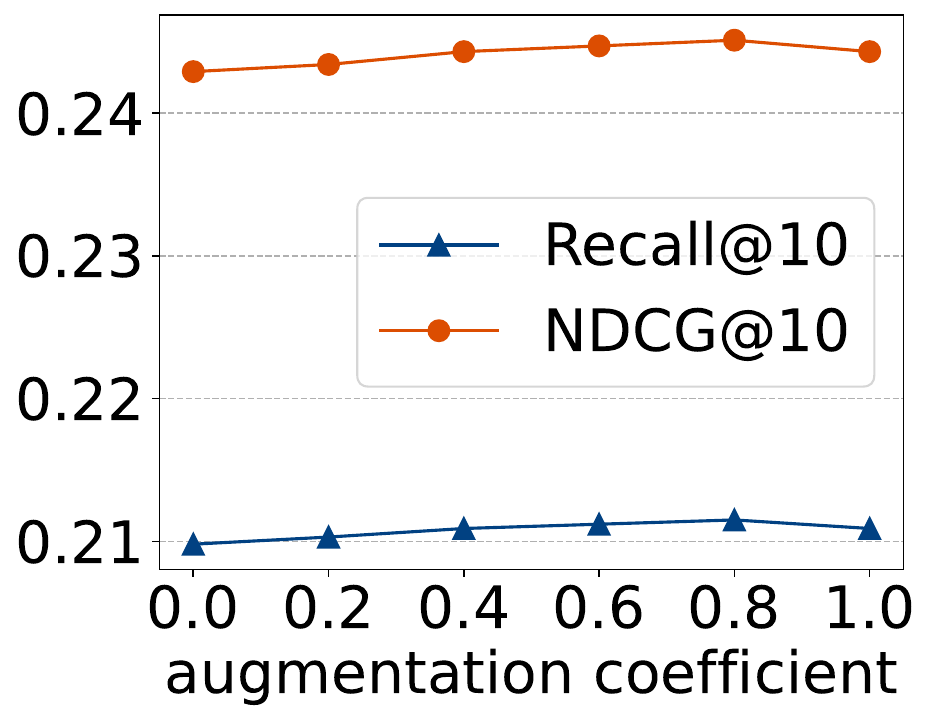}
    \label{fig:yelp-loss}
    \end{minipage}%
    }
    \subfigcapskip=-6pt
    \subfigure[Yelp]{
    \begin{minipage}[c]{0.15\textwidth}
    \centering
    \includegraphics[width=\linewidth]{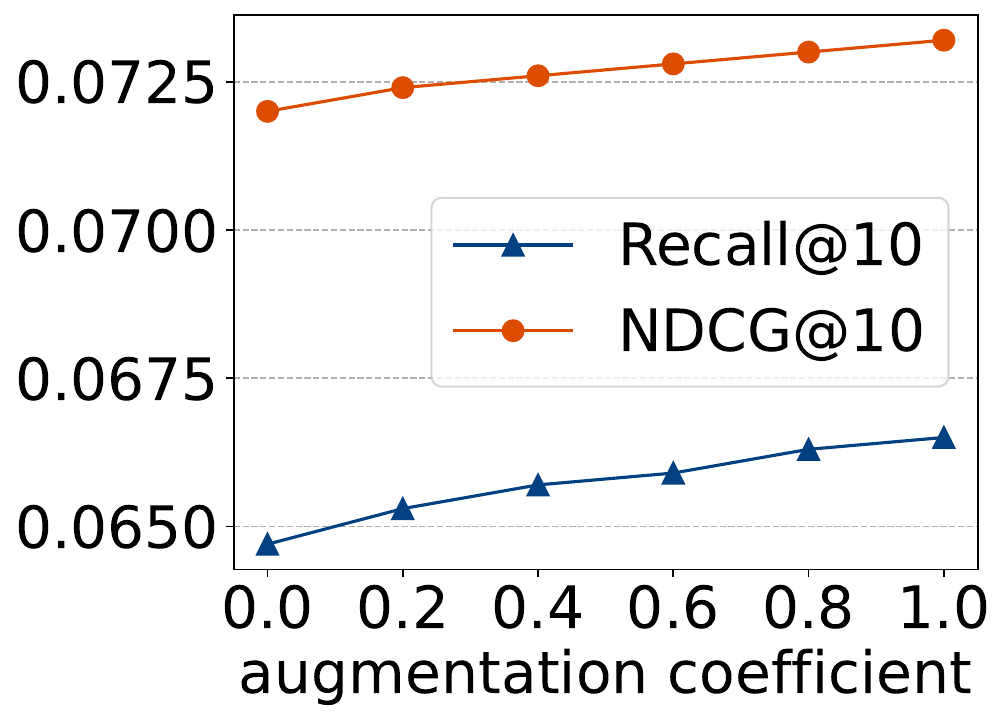}
    \label{fig:amazon-loss}
    \end{minipage}%
    }
    \caption{Impacts of augmentation coefficient.}
    \label{coe}
\end{figure}

\subsection{Model Sensitivity (RQ3)}
\subsubsection{Impacts of the Size of Support Sets}
\label{sec:sparsity}
Interactions are significant data for cold-start recommendation and the sparsity of data influences the quality of interactions. To investigate the effect of support set sparsity (i.e., data sparsity) levels, we range the size on the support set of interacted items from 5 to 70. Since the users that interact with more than 70 are too few, we do not analyze these users. The results are shown in Figure \ref{fig:sparsity}. Overall, all methods perform better with a larger support set. However, when the support set is reduced, the performance degradation on TMAG is the smallest of all approaches. TMAG's recommendation performance is excellent even when the support set is short and performs well regardless of the length of the interaction history. Note that the longer the historical interactions, the smaller the number of users on all datasets. Due to the limited sample size, the performance is thought to be unstable.


\subsubsection{Impacts of Augmentation Coefficient}
We study the performance of interaction augmentation with regard to various coefficients in order to evaluate their robustness. Experiment is also conducted in a well-constrained setting on three datasets by varying $\alpha$ in Equation \eqref{gen} in range of $\{0, 0.2, 0.4, 0.6, 0.8, 1.0\}$. The results are illustrated in Figure \ref{coe}. We can observe that TMAG generalizes well to different coefficients, which demonstrates the effectiveness of the proposed framework. Between the two augmentation strategies, graph augmentation plays a dominant role. The model performs the best on DBook and MovieLens when $\alpha$ is 0.8, while $\alpha$ is 1 on Yelp dataset. This may because that attribute information on yelp is too little, which reduces the credibility of attribute augmentation.

\begin{figure}[t]
    \setlength{\abovecaptionskip}{-0.1cm}
    \setlength{\belowcaptionskip}{-0.1cm}
    \centering
    \centering
    \subfigcapskip=-6pt
    \subfigure[NGCF]{
    \begin{minipage}[c]{0.24\textwidth}
    \centering
    \includegraphics[width=0.9\linewidth]{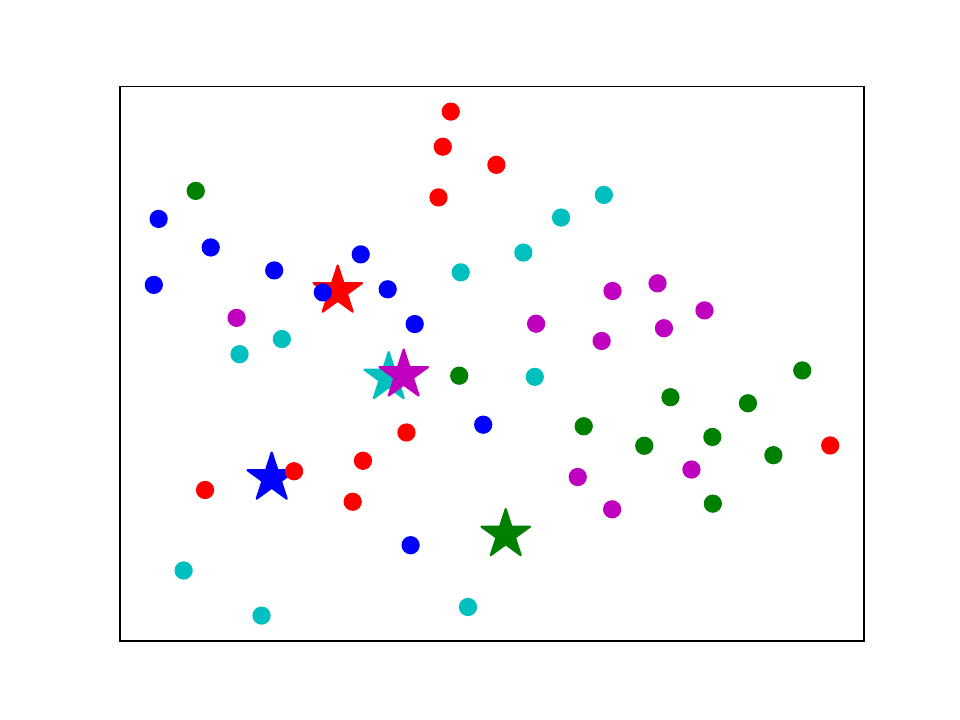}
    \label{fig:tsne_RAE}
    \end{minipage}%
    }\hspace{-5mm}
    \subfigcapskip=-6pt
    \subfigure[TMAG]{
    \begin{minipage}[c]{0.24\textwidth}
    \centering
    \includegraphics[width=0.9\linewidth]{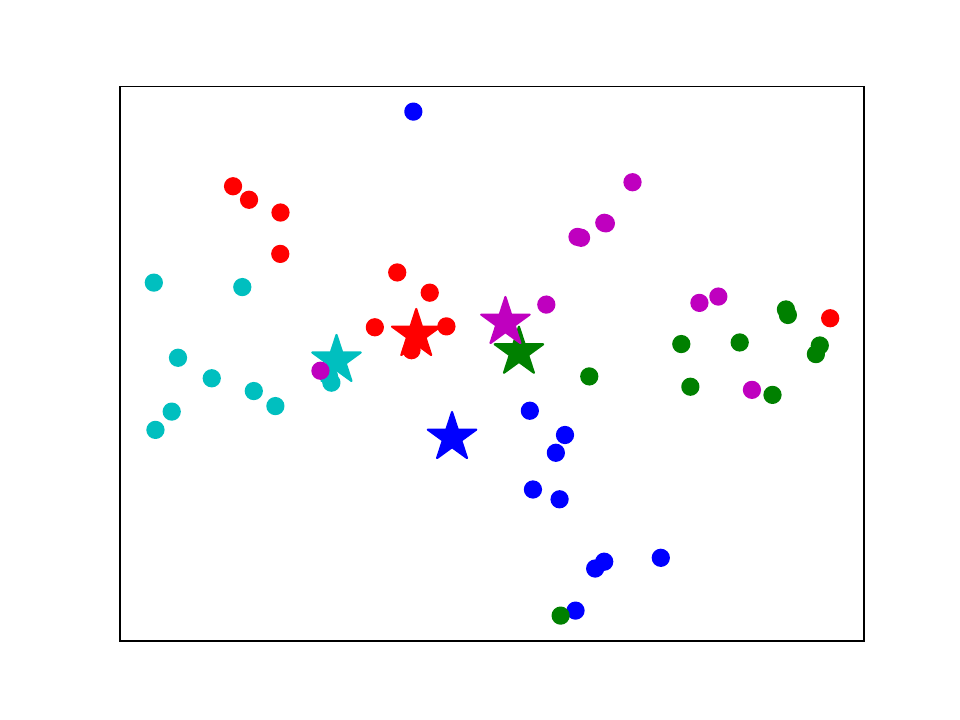}
    \label{fig:tsne_SCVG}
    \end{minipage}%
    }
    \caption{Visualization of t-SNE projected representations derived from NGCF and TMAG. Users are randomly selected in the same task from Yelp and are represented by stars. The points with the same color are users' interacted items.}
    \label{fig:node}
    \vspace{-3mm}
\end{figure}

\subsection{Case Study (RQ4)}
In this section, we will attempt to comprehend how the task alignment promotes the representation learning in the embedding space. Towards this end, we randomly select 5 users in the same task from the Yelp dataset along with their interacted items and visualize the learned user embedding vectors derived from NGCF and TMAG with t-SNE algorithm \cite{van2008visualizing}. Figure \ref{fig:node} provides the results. Note that the items are from the query set in task1, which are not paired with users in the training phase. Compared with NGCF, we can observe that users in the same task tend to be closer and embeddings of items interacted by the same users tend to form the clusters. It indicates that task alignment is capable of mining the latent clustering knowledge to learn more personalized node characteristics.

\section{Conclusion and future work}
In this paper, we propose TMAG to solve the cold-start problem at the model-level and at the feature-level simultaneously. At the model-level, we propose a task aligned constructor to capture the latent clustering knowledge that can be rapidly adapted to new users, which can address the local optimum issues. We also adopt a task-wise attribute contrastive regularization to enhance the latent clustering knowledge. At the feature-level, we combine the graph structure information and attribute information to augment the adjacency matrix of the graph, which alleviates the data sparsity problem. Extensive experiments on three real-world datasets demonstrate the effectiveness of our model for cold-start recommendation. In the future, we would like to expand our work in the following two directions. First, we will extend our model to address cold-start problem from the standpoint of items. Second, we would like to migrate our model to the sequential recommendation scenarios to investigate its effectiveness.

\balance
\bibliographystyle{ACM-Reference-Format}
\bibliography{main}
\balance
\end{document}